\documentclass[twocolumn,secnumarabic,amssymb, nobibnotes, aps, prl,superscriptaddress]{revtex4-2}

\usepackage{etoolbox}
\usepackage{graphicx}  
\usepackage{epstopdf}
\usepackage{dcolumn}   
\usepackage{bm}        
\usepackage{amssymb}   
\usepackage{amsmath}   
\usepackage{amsthm}
\usepackage{chemarrow}
\usepackage{color}
\usepackage{mathrsfs}
\usepackage{float}
\usepackage{bbold}
\usepackage{bbm}

\begin{document}

	
	
	\title{Quantum simulation of a general \MakeLowercase{A}nti-$\mathcal{PT}$-symmetric Hamiltonian with a trapped ion qubit}
	
	\date{\today}

	\author{Ji Bian}
	\thanks{These authors contributed equally to this work.}
	\affiliation{School of Physics and Astronomy,
		Sun Yat-Sen University,
		Zhuhai,
		519082,
		China}
	\affiliation{Center of Quantum Information Technology,
		Shenzhen Research
		Institute of Sun Yat-sen University,
		Shenzhen,
		518087,
		China}
	\author{Pengfei Lu}
	\thanks{These authors contributed equally to this work.}
	\affiliation{School of Physics and Astronomy,
		Sun Yat-Sen University,
		Zhuhai,
		519082,
		China}
	\author{Teng Liu}
	\affiliation{School of Physics and Astronomy,
		Sun Yat-Sen University,
		Zhuhai,
		519082,
		China}
	\author{Hao Wu} 
	\affiliation{School of Physics and Astronomy,
		Sun Yat-Sen University,
		Zhuhai,
		519082,
		China}
	\author{Xinxin Rao}
	\affiliation{School of Physics and Astronomy,
		Sun Yat-Sen University,
		Zhuhai,
		519082,
		China}
	\author{Kunxu Wang}
	\affiliation{School of Physics and Astronomy,
		Sun Yat-Sen University,
		Zhuhai,
		519082,
		China}
	\author{Qifeng Lao}
	\affiliation{School of Physics and Astronomy,
		Sun Yat-Sen University,
		Zhuhai,
		519082,
		China}
	\author{Yang Liu}
	\affiliation{School of Physics and Astronomy,
		Sun Yat-Sen University,
		Zhuhai,
		519082,
		China}
	\affiliation{Center of Quantum Information Technology,
		Shenzhen Research
		Institute of Sun Yat-sen University,
		Shenzhen,
		518087,
		China}
	\author{Feng Zhu}
	\affiliation{School of Physics and Astronomy,
		Sun Yat-Sen University,
		Zhuhai,
		519082,
		China}
	\affiliation{Center of Quantum Information Technology,
		Shenzhen Research
		Institute of Sun Yat-sen University,
		Shenzhen,
		518087,
		China}
	\author{Le Luo}
	\email{luole5@mail.sysu.edu.cn}
	\affiliation{School of Physics and Astronomy,
		Sun Yat-Sen University,
		Zhuhai,
		519082,
		China}
	\affiliation{Center of Quantum Information Technology,
		Shenzhen Research
		Institute of Sun Yat-sen University,
		Shenzhen,
		518087,
		China}
	\affiliation{
		State Key Laboratory of Optoelectronic
		Materials and Technologies,
		Sun Yat-Sen
		University (Guangzhou Campus),
		Guangzhou,
		510275,
		China}

\begin{abstract}
Non-Hermitian systems satisfying parity-time ($\mathcal{PT}$) symmetry have
aroused considerable interest owing to their exotic
features. Anti-$\mathcal{PT}$ symmetry is an important counterpart of the $\mathcal{PT}$ symmetry, and has been studied in various classical systems.
Although  a
 Hamiltonian with anti-$\mathcal{PT}$ symmetry only differs from its $\mathcal{PT}$-symmetric counterpart in a global $\pm i$ phase, the information and energy exchange between systems and  environment are different under them.
 It is also suggested theoretically that anti-$\mathcal{PT}$ symmetry is a useful concept in the context of quantum information storage with qubits coupled to a bosonic bath.
So far, the observation of anti-$\mathcal{PT}$
symmetry in individual quantum systems remains elusive. Here, we
implement an anti-$\mathcal{PT}$-symmetric Hamiltonian of a single qubit in a single
trapped ion by a designed microwave and optical control-pulse sequence. We
characterize the anti-$\mathcal{PT}$ phase transition by mapping out the eigenvalues at different ratios between coupling strengths and dissipation rates. The full information of the quantum state is also obtained by quantum state tomography. Our work allows quantum simulation of genuine open-system feature of an anti-$\mathcal{PT}$-symmetric system, which
paves the way for utilizing non-Hermitian properties for quantum information processing.

\end{abstract}



\maketitle


\section{Introduction}
In quantum mechanics, the Hamiltonian of a system is typically taken to be Hermitian in order to generate real eigenenergy spectrum. However, as pointed out by Bender et al.
\cite{bender1999pt}, non-Hermitian Hamiltonians
obeying parity-time ($\mathcal{PT}$) symmetry could still give real
eigenenergies. $\mathcal{PT}$-symmetric Hamiltonians exhibit various exotic
behaviors, one of which is $\mathcal{PT}$-symmetry-breaking
transitions that show up at an exceptional point (EP). At an EP, the eigenenergies and eigenstates of the Hamiltonian
become degenerate
\cite{bender2007faster,zheng2013observation,li2019observation,wang2021observation,ding2021experimental,wu2019observation}.
Experimental studies on $\mathcal{PT}$-symmetry in classical systems have stimulated various applications such as lasing
\cite{hodaei2014parity}, optimal energy transfer
\cite{doppler2016dynamically} and enhanced sensing
\cite{hodaei2017enhanced}, etc. 
Recently, $\mathcal{PT}$-symmetric Hamiltonians are
also constructed in genuine quantum systems, e.g., ultracold atoms
\cite{li2019observation}, NV-centers \cite{wu2019observation},
trapped ions \cite{wang2021observation,ding2021experimental}, and
superconducting quantum circuits \cite{naghiloo2019quantum}, etc. These
allow quantum signatures such as perfect quantum coherence at EP to
be revealed \cite{wang2021observation}. Moreover, the topological
structure of exceptional points is utilized to realize, e.g., robust
quantum control \cite{liu2021dynamically}.

Not limited to $\mathcal{PT}$ symmetry, EP also appear in 
systems with anti-$\mathcal{PT}$ symmetry \cite{choi2018observation,zheng2020efficient,zhang2022anti}.
As an important counterpart of the $\mathcal{PT}$ symmetry, anti-$\mathcal{PT}$ symmetry 
has been studied in various physical systems \cite{li2019anti,nair2021enhanced}, including coupled waveguides \cite{konotop2018odd},
nanophotonics \cite{fan2020antiparity}, microcavities \cite{zhang2020synthetic}, lossy resonators \cite{zhang2020breaking}, optical fibres \cite{bergman2021observation},
optical systems with atomic media \cite{jiang2019anti,peng2016anti},
and electrical circuit resonators \cite{choi2018observation}. An
anti-$\mathcal{PT}$-symmetric Hamiltonian $H_{\textrm{APT}}$ satisfying
$\{\textrm{PT},H_{\textrm{APT}}\}=0$ is related to a $\mathcal{PT}$-symmetric Hamiltonian $H_{\textrm{PT}}$ by
$H_{\textrm{APT}}=\pm i H_{\textrm{PT}}$. Here $\textrm{P}$ and
$\textrm{T}$ denote parity and time reflection operation,
$\{\cdot\}$ denotes anticommutator \cite{bender1999pt}.
As a result, properties similar to $\mathcal{PT}$-symmetric Hamiltonians such as eigenstates coalescing at EP and spontaneous symmetry-breaking transition show up \cite{choi2018observation}. 
The information and energy exchange between  anti-$\mathcal{PT}$-symmetric systems and environment are different from the $\mathcal{PT}$-symmetric counterpart.
 These result in different information-exchange scheme, e.g., coherence flow \cite{fang2021experimental}, between anti-$\mathcal{PT}$-symmetric systems and the $\mathcal{PT}$-symmetric counterpart. 
Interestingly, recent
theoretical works \cite{gardas2016pt,cen2022anti} studied the evolution of
qubits under $H_{\textrm{APT}}$ when coupled to a bosonic bath, and claim that these qubits decohere more slowly compared to those under Hermitian or $\mathcal{PT}$-symmetric Hamiltonians. This suggests that the anti-$\mathcal{PT}$-symmetric Hamiltonian as a whole is a useful concept in advancing quantum
information processing under decohering environment. 
Therefore, it is important to carry out further experimental research on the eigensystem structure and environmental energy-exchange scheme of a generic anti-$\mathcal{PT}$-symmetric system.

So far the experiments on anti-$\mathcal{PT}$-symmetric Hamiltonian construction are
mainly limited to classical systems
\cite{choi2018observation} and ensemble spin systems \cite{wen2020observation}. Here, we demonstrate
the implementation of an individual quantum system acquiring
anti-$\mathcal{PT}$ symmetry. We realize the evolution under $H_{\textrm{APT}}$
by designing appropriate pulse sequences, as a standard technique in
quantum simulation experiments \cite{georgescu2014quantum}. The
sequence consists of an evolution under a passive $\mathcal{PT}$-symmetric
Hamiltonian $H_{\textrm{M}}$ sandwiching between two $\pi/2$ pulses
with opposite phases. The evolution under $H_{\textrm{M
}}$ is
achieved by a dissipation scheme as  demonstrated in
\cite{li2019observation,wang2021observation,ding2021experimental}, in the context of
$\mathcal{PT}$-symmetric Hamiltonian construction. We experimentally verify the
anti-$\mathcal{PT}$-symmetric Hamiltonian by studying its anti-$\mathcal{PT}$ phase transition
behavior. We obtain the eigenvalues at different coupling strength by preparing a certain initial state, evolving it under $H_{\textrm{APT}}$ for some known duration, and measure the overlap between the evolved state and the initial state, similar to \cite{ding2021experimental}.  The results clearly show the transition
from anti-$\mathcal{PT}$ symmetry region to anti-$\mathcal{PT}$ symmetry broken region. The full information of the quantum states, i.e., the population as well as the coherence,  is also obtained by quantum state tomography. This
enables further studies on anti-$\mathcal{PT}$-symmetric systems, e.g., the information flow
\cite{wen2020observation} and the topological state transfer near an
EP \cite{zhang2019dynamically} to be conducted. Our work could also
serve as a first step towards harnessing non-Hermitian $\mathcal{PT}$ or anti-$\mathcal{PT}$
physics to advance the field of quantum information processing
\cite{naghiloo2019quantum,chang2020second}.

\section{Anti-$\mathcal{PT}$-symmetric Hamiltonian construction}
The anti-$\mathcal{PT}$-symmetric Hamiltonian we want to construct reads
\begin{equation}\label{h1}
H_{\textrm{APT}}=-2J I_z+2i\Gamma I_x-i\Gamma\mathbf{I},
\end{equation}
where
\begin{equation*}
I_x= \frac{1}{2}\begin{pmatrix} 0 & 1 \cr 1 & 0 \end{pmatrix}, \
I_y= \frac{1}{2}\begin{pmatrix} 0 & -i \cr i & 0 \end{pmatrix}, \
I_z= \frac{1}{2}\begin{pmatrix} 1 & 0 \cr 0 & -1 \end{pmatrix}
\end{equation*}
are angular momentum operators, and $\Gamma, J$ are real
parameters. Indeed, $\{\textrm{PT},H_{\textrm{APT}}\}=0$, satsifying
the anti-$\mathcal{PT}$ requirement \cite{peng2016anti}, where $\textrm{P}=2I_x$,
$\textrm{T}=*$ denotes complex conjugation operation. Given a qubit
with eigenstates $|0\rangle=(1,0)^{\textrm{T}}$ and
$|1\rangle=(0,1)^{\textrm{T}}$, $H_{\textrm{APT}}$ in Eq. \eqref{h1}
could be realized as follows.

\begin{figure}[htb]  
	\makeatletter
	\def\@captype{figure}
	\makeatother
	\includegraphics[scale=0.91
	]{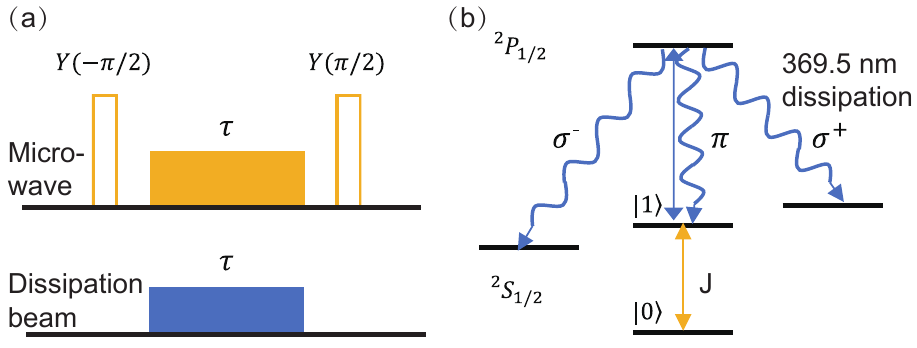}
	\caption{(color online). (a) The pulse sequence utilizng microwave and dissipative beam to realize $H_{\textrm{APT}}$. $Y(\pm \pi/2)$ denotes rotation along $\pm y$ axis by $\pi/2$. $\tau$ is the duration of the middle pulse which is realized by simultaneous application of microwave and dissipative beam.  (b) The trapped $^{171}$Yb$^{+}$ ion as a qubit.  The hyperfine states in $^{2}$S$_{1/2}$ with $m=0$ correspond to qubit levels $|0\rangle$ and $|1\rangle$. The $369.5$ nm dissipative beam, the spontaneous decay channel, and the microwave control field with coupling strength $J$ are explained in the main text.}
	\label{fig01}
\end{figure}

\textit{1)} We first realize a passive $\mathcal{PT}$-symmetirc Hamiltonian
through a spin-dependent dissipation scheme, which is first realized in cold atoms \cite{li2019observation}, then in trapped ions \cite{wang2021observation}. This is
achieved by coupling $|0\rangle$ and $|1\rangle$ by a control field (e.g., a microwave field)
with coupling strength $J$, and generate a loss of population on
$|1\rangle$ with effective loss rate $4\Gamma$. The population loss
could be achieved by adding a dissipative beam
\cite{ding2021experimental} as will be explained in the following section.
Now we have constructed the passive $\mathcal{PT}$-symmetric Hamiltonian
$H_{\textrm{M}}=2i\Gamma I_z+ 2JI_x-i\Gamma\mathbf{I}$.
\textit{2)} We then realize the final anti-$\mathcal{PT}$-symmetric Hamiltonian
\eqref{h1} by designing appropriate pulse sequences, as is typically
done in quantum simulation experiments \cite{georgescu2014quantum}.
According to the identities 
\begin{equation}
Re^{A}R^{\dagger}=e^{RAR^{\dagger}}, \, \textrm{if} \, RR^{\dagger}=\mathbf{I},
\end{equation}
for arbitrary square matrices $A$; and
\begin{equation}
e^{-i\theta I_{\alpha}}I_{\beta}e^{i\theta I_{\alpha}}=I_{\beta}\cos\theta+I_{\gamma}\sin\theta,
\end{equation}
where $\{\alpha, \beta, \gamma\}$ are cyclic permutations of \{$x, y, z$\},
the evolution of the system under $\eqref{h1}$, $e^{-iH_{\textrm{APT}}\tau}$, could be realized by sandwiching the evolution under $H_{\textrm{M}}$ between two $\pi/2$ pulses along $\pm y$ axis. That is, denote $R_y(\theta)=e^{-i\theta I_y}$ the rotation along $y$ for $\theta$, then by taking $R=R_y(\pi/2)$ and $A=-iH_{\textrm{M}}\tau$, we have
\begin{equation*}
\begin{aligned}
R_y(\frac{\pi}{2})e^{-iH_{\textrm{M}}\tau}R_y(-\frac{\pi}{2})&=e^{-iR_y(\frac{\pi}{2})H_{\textrm{M}}R_y(-\frac{\pi}{2})\tau}\\
&=e^{-i(2i\Gamma I_x-2JI_z-i\Gamma \mathbf{I})\tau}=e^{-iH_{\textrm{APT}}\tau}.
\end{aligned}
\end{equation*}
The pulse sequence is illustrated in Fig. \ref{fig01}, (a).

Finally, we obtain the evolution under anti-$\mathcal{PT}$-symmetric Hamiltonian for
time $\tau$. $\tau$ equals the evolution time under
$H_{\textrm{M}}$, i.e., the duration of the middle pulse in Fig.
\ref{fig01}, (a). One could change the duration of the middle pulse
to obtain evolutions under $H_{\textrm{APT}}$ for arbitrary times.
It is noted that, a Floquet scheme \cite{ding2021experimental,ding2021information} has used similar pulse sequence, but needs many cycles to generate the desired evolutions. 

\begin{figure}[htb]  
	\centering
	\makeatletter
	\def\@captype{figure}
	\makeatother
	\includegraphics[scale=0.85
	]{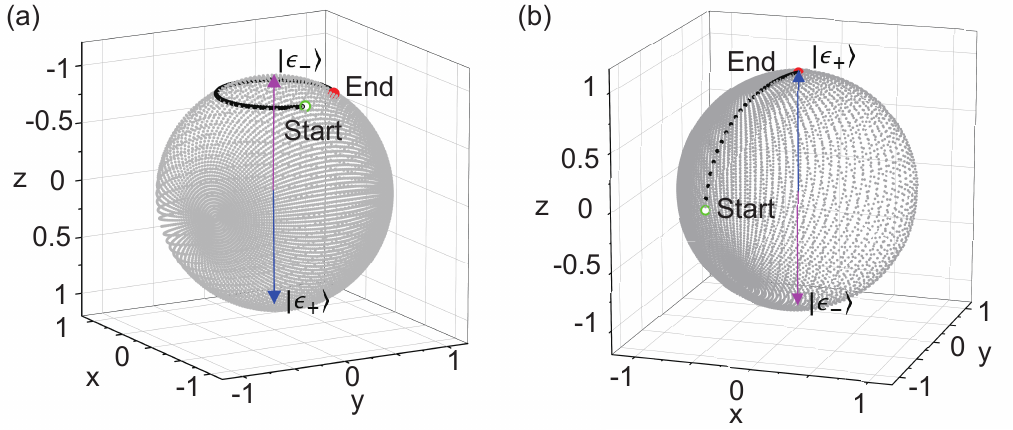}
	\caption{(color online). Evolution of the state under $H_{\textrm{M}}$ on the non-Hermitian Bloch sphere. The choice of coordinates are given in the main text. $|\epsilon_{\pm}\rangle$ are marked on the sphere. The state starts from $\frac{1}{\sqrt{2}}(|0\rangle-|1\rangle)$, and evolves under $H_{M}$ for $\tau$. Here  $J=0.06$ MHz , $\tau=50$ $\mu$s, $\Gamma=0.03$ MHz in (a) and $\Gamma=0.12$ MHz in (b).}
	\label{fig0s}
\end{figure}

Starting from $|0\rangle$, without the dissipative beam, the state is fixed during the middle pulse and will return to $|0\rangle$ at the end of the sequence. The addition of dissipation during the middle pulse evolves the state to a different point in the Hilbert space, which essentially result in the evolution of $H_{\textrm{APT}}$. To better understand the evolution of the quantum state in the Hilbert space under dissipation, we construct a non-Hermitian Bloch sphere using CPT symmetry \cite{bender2005introduction} as illustrated in Fig. \ref{fig0s}. The system is evolving under $H_\textrm{M}$ during the middle pulse, for such a passive $\mathcal{PT}$-symmetric non-Hermitian Hamiltonian, a linear operator C exists, which satisfies $[H_{\textrm{M}},\textrm{C}]=0$ and $[\textrm{PT},\textrm{C}]=0$ \cite{bender2002complex,bender2005introduction}. Replacing the complex conjugate with the CPT-conjugate, the new inner product for an arbitrary state $|\psi\rangle$ can be written as the form of Dirac inner product \cite{bender2004scalar}:
$
\langle \psi|\psi\rangle^{\textrm{CPT}}=\langle \psi|\textrm{P}^{T}\textrm{C}^{T}|\psi\rangle.
$
The structure of the resulting new Hilbert space (CPT-inner-product space) depends on C, which could be written as $\textrm{C}=2/\sqrt{1-r^2}(I_x+irI_z)$, where $r:=\Gamma/J$.
Then, taking the normalized eigenstates $|\epsilon_{+}\rangle$ and $|\epsilon_{-}\rangle$ of $H_{\textrm{M}}$ as a basis, an arbitrary state $|\psi\rangle$ could be written as 
$
|\psi\rangle^{\textrm{CPT}}=R\cos{\frac{\Theta}{2}}|\epsilon_{+}\rangle+R\sin{\frac{\Theta}{2}}e^{i\Phi}|\epsilon_{-}\rangle.
$
As $\langle \epsilon_{\pm}|\epsilon_{\pm}\rangle ^{\textrm{CPT}}=1$ and $\langle \epsilon_{\pm}|\epsilon_{\mp}\rangle ^{\textrm{CPT}}=0$, the evolution of $|\psi\rangle$ under $H_{\textrm{M}}$ could be demonstrated on a newly constructed non-Hermitian Bloch sphere with radius $R$. The axes are chosen such that the point $(x=0,y=0,z=\pm R)$ represents $R|\epsilon_{\pm}\rangle$, $\Theta$ equals the angle spanned by the state vector and $z$ axis, and $\Phi$ equals the angle between the state vector and $x$ axis. As an example, the trajectory under $H_{\textrm{M}}$ with initial state $|\psi_0\rangle=1/\sqrt{2}(|0\rangle-|1\rangle)$, $J=0.06$ MHz,  $\tau=50$ $\mu$s, $\Gamma=0.03$ MHz is plotted on the unit ($R=1$) non-Hermitian Bloch sphere, as illustrated in Fig. \ref{fig0s}, (a). If $\Gamma$ is changed to $0.12$ MHz, while other parameters remain unchanged, the trajectory is plotted on another non-Hermitian Bloch sphere, as shown in Fig. \ref{fig0s}, (b). If $\Gamma=0$, the state is fixed on the sphere. As $\Gamma$ gets larger, the state starts to evolve and trajectories appear on the corresponding CPT sphere. This helps to better understand the evolution under both Hermitian coupling and non-Hermitian dissipation.

\section{Experimental implementation and verification of the anti-$\mathcal{PT}$-symmetric Hamiltonian}\label{exp}

We conduct the experiment on a trapped $^{171}$Yb$^{+}$ quantum
processer. Qubit levels $|0\rangle$ and $|1\rangle$ correspond to the two hyperfine ground states $|F=0,m=0\rangle$ and
$|F=1,m=0\rangle$, as
demonstrated in Fig. \ref{fig01}, (b). The
qubit splitting $\omega_{\textrm{HF}}\approx 12.6$ GHz. The
two levels are coupled by microwave fields with coupling strength $J$. The population loss is realized by a dissipation
scheme, as demonstrated in
\cite{ding2021experimental}.  The ion is excited  from $|F=1,m=0\rangle$ to 
$^2$P$_{1/2}$ state by a $369.5$ nm
dissipative beam, which contains only $\pi$-polarization
components. This ensures that excitations from 
states $|F=1,m=\pm 1\rangle$ are forbidden accroding to selection rules. Through spontaneously emitting $\sigma$ or $\pi$ polarized photons, the
excited $^2$P$_{1/2}$ state will decay to the $|F=1,m=0,\pm 1\rangle$  states in the $^2$S$_{1/2}$ manifold. This effectively generates a population loss on  $|1\rangle$
at a loss rate $4\Gamma$ \cite{ding2021experimental}. The whole scheme is
illustrated in Fig. \ref{fig01}, (b). 

\begin{figure}[h]  
	\centering
	\makeatletter
	\def\@captype{figure}
	\makeatother
	\includegraphics[scale=1]{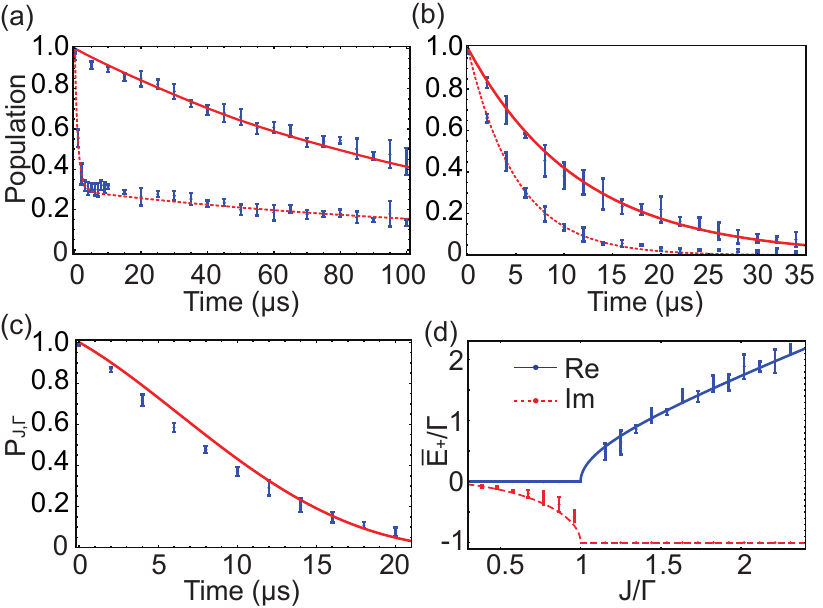}
	\caption{(color online). (a) shows population on $|0\rangle$ as a function of evolution time, with $J=0.06$ MHz. Blue dots are experimental data, and the red curve is a fit to $\rho_{00}$ in \eqref{p0}. $\Gamma$ is fitted to be $0.004$ MHz for the solid curve and $0.53$ MHz for the dashed curve. Hence the anti-$\mathcal{PT}$-symmetry is broken (preserved) for the solid curve (dashed curve). 
		(b) shows the evolution of $\rho_{11}$ under pure dissipation. $\Gamma$ is fitted to be 0.022 (0.050) for the solid (dashed) curve.
		(c) shows the theoretical $P_{J,\Gamma}(\tau)$ (solid lines) and the measured $P_{J,\Gamma}(\tau)$ (dots), $\Gamma=0.022$ MHz and $J=0.065$ MHz. 
		(d) shows the eigenvalue $\bar{E}_{+}$. Dots representing average value of the eigenvalue, together with the error bars, are obtained by repeating the experiment $3$ times as explained in the main text. The solid (dashed) curve represents real and imaginary part of the theoretical eigenvalue.}
	\label{fig02}
\end{figure}

\begin{figure}[h]  
	\centering
	\makeatletter
	\def\@captype{figure}
	\makeatother
	\includegraphics[scale=1]{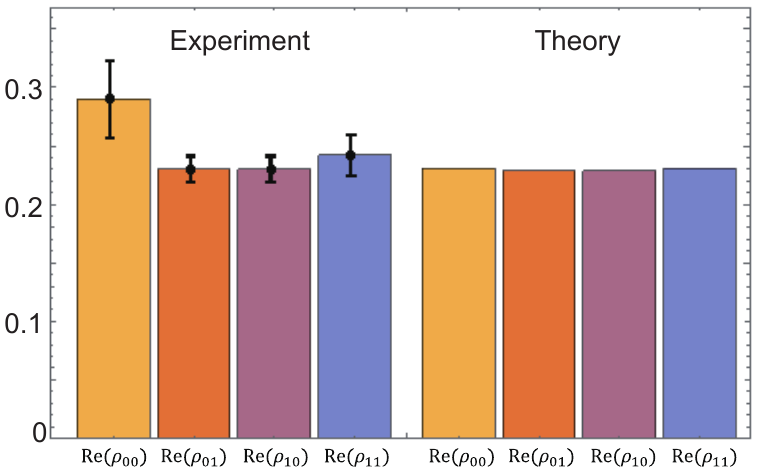}
	\caption{(color online). Tomography of the density matrix $\rho$ after an evolution under $H_{\textrm{APT}}$ for $10$ $\mu s$, with $J/\Gamma \approx 0.15$. The imaginary parts of the density matrix are close to $0$ and are not shown.}
	\label{fig03}
\end{figure}

The evolution under $H_{\textrm{APT}}$ is realized by the sequence
depicted in Fig.\ref{fig01}, (a): 1) Apply a $\pi/2$ microwave pulse
along $-y$ axis in the rotating frame. 2) Apply a microwave field
along $x$ axis with strength $J$ and a dissipative beam with
dissipation rate $4\Gamma$ simultaneously, for a duration $\tau$.
This creates an evolution under $H_{\textrm{M}}$ for $\tau$. 3)
Finally, a $\pi/2$ pulse along $y$ is implemented. The resulting
whole evolution is $U(\tau)=e^{-iH_{\textrm{APT}}\tau}$, i.e., an
evolution under $H_{\textrm{APT}}$ for time $\tau$.
Denote the initial state $|\Psi(0)\rangle$, and the density matrix at time $\tau$, $\rho(\tau)=|\Psi(\tau)\rangle \langle \Psi(\tau)|$, where $|\Psi(\tau)\rangle=e^{-iH_{\textrm{APT}}\tau}|\Psi(0)\rangle$. 
Starting from $|0\rangle$, the density matrix elements $\rho_{ij}(\tau)=\langle i|\rho(\tau)|j\rangle$ read
\begin{small}
\begin{equation}\label{p0}
\begin{aligned}
\rho_{00}(\tau)=&e^{-2\Gamma
	\tau}\{\cos^2(\tau\sqrt{J^2-\Gamma^2})+[\frac{J\sin(\tau\sqrt{J^2-\Gamma^2})}{\sqrt{J^2-\Gamma^2}}]^2\}.\\
\rho_{11}(\tau)=&\frac{e^{-2\Gamma \tau}\Gamma^2\sin^2(\sqrt{J^2-\Gamma^2}\tau)}{J^2-\Gamma^2}.\\
\rho_{01}(\tau)=&e^{-2\Gamma \tau}\Gamma \frac{i\sin (\sqrt{J^2-\Gamma^2}\tau)}{\sqrt{\Gamma^2-J^2}}[\cos(\sqrt{J^2-\Gamma^2}\tau)\\
&+\frac{J\sin(\sqrt{J^2-\Gamma^2}\tau)}{\sqrt{\Gamma^2-J^2}}].
\end{aligned}
\end{equation}
\end{small}
$\rho_{00}(\tau)$ is measured for both $J/\Gamma <1$ and $J/\Gamma >1$, as shown in Fig. \ref{fig02}, (a). $J=0.06$ MHz is obtained separately by fitting the Rabi oscillation. $\Gamma$ could be obtained by fitting according to $\rho_{00}(\tau)$ in \eqref{p0}, e.g., $\Gamma=0.004$ $(0.53)$ MHz for the solid (dashed) curve. Similarly, $\rho_{11}(\tau)$ and $\rho_{01}(\tau)$ could be obtained by standard quantum state tomography technique. $\Gamma$ could also be independently calibrated by preparing the system in $|1\rangle$, let it evolve under pure dissipation (the corresponding Hamiltonian is $-2i\Gamma |1\rangle \langle 1 |$) for $\tau$, measure the final population on $|1\rangle$ and fit it with $e^{-4\Gamma \tau}$. For example, in Fig. \ref{fig02}, (b), two such processes are plotted and $\Gamma$ is fitted to be $0.022$ $(0.050)$ MHz for the solid (dashed) curve.  

To verify the constructed anti-$\mathcal{PT}$-symmetric system, we experimentally study the
anti-$\mathcal{PT}$ phase transition behavior. The eigenvalues $E_{\pm}$ of the anti-$\mathcal{PT}$-symmetric Hamiltonian \eqref{h1}, or equivalently $\bar{E}_{\pm}=E_{\pm}/\Gamma$, could serve to characterize this transition, 
\begin{equation}\label{eigenvalue}
\bar{E}_{\pm}=-i \pm \frac{\sqrt{J^2-\Gamma^2}}{\Gamma}.
\end{equation}
One way \cite{wu2019observation} to obtain $\bar{E}_{\pm}$ and verify the constructed system would be calibrating $J$ and $\Gamma$ independently, and curve fitting $\rho_{00}(\tau)$ to the theoretical predictions. Here we utilize a more effective method to measure the overlap between a certain initial state and the corresponding evolved state \cite{ding2021experimental}, as 
\begin{equation}\label{pjg}
\begin{aligned}
P_{J,\Gamma}(\tau)&=|\frac{\langle 0|+i\langle 1|}{\sqrt{2}}\textrm{exp}(-iH_{\textrm{APT}}\tau)\frac{|0 \rangle-i|1\rangle}{\sqrt{2}}|^2 \\
&=\cos^2(\sqrt{J^2-\Gamma^2}\tau)e^{-2\Gamma \tau},
\end{aligned}
\end{equation}
one could first measure $P_{J,\Gamma}(\tau)$ by preparing the initial state $(|0\rangle-i|1\rangle)/\sqrt{2}$, let it evolve under $H_{\textrm{APT}}$ for a known $\tau$, and measure the overlap between the evolved state and $(|0\rangle-i|1\rangle)/\sqrt{2}$ (left multiply its conjugate transpose). Together with the independently measured $\Gamma$, one could obtain $\sqrt{J^2-\Gamma^2}$ by calculating the arccos or arccosh function followed by a division over $\tau$, hence $\bar{E}_{\pm}$ is also obtained.

$P_{J,\Gamma}(\tau)$ could be measured by starting from $|0\rangle$, apply a $R_x(\pi/2)$ pulse, let the system evolve according to $e^{-iH_{\textrm{APT}}\tau}$, then apply a $R_x(-\pi/2)$ pulse and readout the population on $|0\rangle$.   For example, in Fig. \ref{fig02}, (c), we first set $\Gamma=0.022$ MHz by an independent calibration (Fig.2, (b), solid curve), and $J=0.065$ MHz, this gives the theoretical curve; then $P_{J,\Gamma}(\tau)$ is measured by the above sequence. The theoretical curve and experimental data are close to each other.
The experimental procedure to obtain $E_{\pm}$ at different $J/\Gamma$ is listed below.

1) $\Gamma$ and $J$ are first calibrated through independent measurements. This gives us a correspondence between the dissipative beam strength (microwave field strength) and $\Gamma$ ($J$). In the following we set the dissipative beam strength corresponding to $\Gamma=0.050$ MHz (Fig.2, (b), dashed curve), and vary the microwave field strength .  

2) The initial state is prepared in $|0\rangle$ by optical pumping. Set the microwave field strength corresponding to $J$.

3) $P_{J,\Gamma}$ is measured at a certain time $\tau_0$, here we choose  $\tau_0=1/J$. At this step, we also separately measure $\Gamma$ as in Fig.\ref{fig02}, (b). 

4) Plug $P_{J,\Gamma}$, $\tau_0$ and $\Gamma$ [obtained in 3)] into \eqref{pjg}, we obtain $\sqrt{J^2-\Gamma^2}$. Plug $\sqrt{J^2-\Gamma^2}$ and $\Gamma$ [obtained in 3)] into \eqref{eigenvalue}, we obtain $\bar{E}_{\pm}$. 

5) Repeat steps 2) to 4) with varying $J$, we obtain $\bar{E}_{\pm}$ at different $J/\Gamma$. Here $J/\Gamma$, i.e., the horizontal axis of Fig. \ref{fig02}, (d), are evaluated by the calibrated values in step 1). 

6) Repeat steps 2) to 5) three times to obtain the average and variance of $\bar{E}_{\pm}$. The result is shown in Fig. \ref{fig02}, (d), where only $\bar{E}_{+}/\Gamma$ is plotted. The error bars reflect the fact that there are noises in $J$, $\Gamma$, the constructed evolution under $H_{\textrm{APT}}$, and the readout process.
%
%

The system stays in the anti-$\mathcal{PT}$ symmetry  region with purely imaginary eigenvalues when $J/\Gamma < 1$. At EP ($J/\Gamma=1$), the
eigenvalues  become degenerate and equals to $-i\Gamma$. The system enters
the anti-$\mathcal{PT}$ symmetry broken region when $J/\Gamma > 1$, and the eigenvalues start to have real components. The experimental results agree well with
the theoretical calculations, as illustrated in Fig. \ref{fig02}, (d).

The full density matrix of the system is further obtained through quantum
state tomography. As an example, Fig. \ref{fig03} shows the
experimental density matrix $\rho_{\textrm{exp}}$ at $10$ $\mu s$, $J/\Gamma \approx 0.15$,
together with the theoretical values $\rho_{\textrm{th}}$. Applying the state fidelity formula $F=|\textrm{Tr}(\bar{\rho}_{\textrm{th}}\bar{\rho}_{\textrm{exp}})|/\sqrt{\textrm{Tr}(\bar{\rho}^2_{\textrm{th}})\textrm{Tr}(\bar{\rho}^2_{\textrm{exp}})}$, where $\bar{\rho}_{\textrm{exp}}$=$\rho_{\textrm{exp}}$/\textrm{Tr}($\rho_{\textrm{exp}}$), $\bar{\rho}_{\textrm{th}}$=$\rho_{\textrm{th}}$/\textrm{Tr}($\rho_{\textrm{th}}$) are experimental and theoretical normalized density matrix \cite{wen2020observation,suter2016colloquium}, the fidelity is  calculated to be $97.3 \% \pm 1.1 \%$.
Simulation results suggest that the errors might be caused by noises in the dissipative beam and readout pulses. The experimental results match theoretical predictions, which
demonstrates the reliability of the constructed anti-$\mathcal{PT}$-symmetric system.

While the passive $\mathcal{PT}$-symmetric Hamiltonian is constructed by the state-dependent dissipation scheme in trapped-ions \cite{wang2021observation,ding2021experimental}, the anti-$\mathcal{PT}$-symmetric one has not been  directly implemented, due to the difficulties in constructing controllable dissipative-coupling between the two qubit states. Our method effectively implements the anti-$\mathcal{PT}$-symmetric Hamiltonian through adding two additional pulses, arbitrary evolutions under $H_{\textrm{APT}}$ could thus be simulated. Note that in the experiment, the constructed Hamiltonian is exactly \eqref{h1}, which is anti-$\mathcal{PT}$-symmetric, there is no need to remove the $i\Gamma\mathbf{I}$ term artificially as is done when simulating $\mathcal{PT}$-symmetric systems \cite{li2019observation,wang2021observation,ding2021experimental}. Although $H_{\textrm{APT}}$ and $H_\textrm{PT}=\pm iH_{\textrm{APT}}$ have similar eigensystem structures, a recent work \cite{fang2021experimental} suggests that they behave differently in the context of coherence flow between the system and environment. When taking the phonon degrees of freedom into account, evolutions under $H_\textrm{APT}$ might decohere more slowly than those under $H_\textrm{PT}$, as studied in \cite{cen2022anti}. This suggests that anti-$\mathcal{PT}$- symmetry is a useful concept in trapped-ion quantum information processing.

~\\
\section{Conclusion}
To conclude, we have successfully implemented an anti-$\mathcal{PT}$-symmetric 
 quantum system  by a single $^{171}$Yb$^{+}$
ion. By sandwiching an evolution under a passive $\mathcal{PT}$-symmetric
Hamiltonian between two $\pi/2$ pulses with opposite phases, we
realize the desired evolution under anti-$\mathcal{PT}$-symmetric Hamiltonian. We
experimentally verify the anti-$\mathcal{PT}$-symmetric Hamiltonian by studying its
anti-$\mathcal{PT}$-symmetric phase transition behaviour. By preparing a certain initial state, evolving it under $H_{\textrm{APT}}$, and measuring the overlap between the final state and the initial state, we
obtain the eigenvalues at different $J/\Gamma$.  The
transition from anti-$\mathcal{PT}$ symmetry region to anti-$\mathcal{PT}$ symmetry broken region,
together with the eigenvalue coalescing, are clearly revealed from
the data. The full information of the quantum states, i.e., the population as well as the coherence, are also obtained by
quantum state tomography. The experimental results agree well with
theoretical predictions.

Based on the constructed anti-$\mathcal{PT}$-symmetric Hamiltonian and the versatile
quantum-control toolbox trapped ions offer \cite{ai2021experimental,wu2021quantum}, further experimental
studies on non-Hermitian physics could be envisioned. For example,
the information retrieval
\cite{ding2021information,wen2020observation} and topological state
transfer \cite{liu2021dynamically,doppler2016dynamically} in
non-Hermitian quantum systems. Our work also paves the way for
harnessing non-Hermitian physics in quantum information-processing
applications, e.g., qubits with non-Hermitian $\mathcal{PT}$ or anti-$\mathcal{PT}$-symmetric
Hamiltonians could have superior coherence times compared to Hermitian
qubits when coupled to a bosonic bath \cite{cen2022anti,gardas2016pt}.

Note that after this work is finished, we become aware of a similar experiment done
recently by L. Ding et al \cite{ding2021information}. They implement a Floquet Hamiltonian requiring multi-cycles of pulses to create anti-$\mathcal{PT}$- symmetric Hamiltonian by periodically driving a dissipative  single trapped ion.

\section*{Declaration of Competing Interest}
The authors declare that they have no conflict of interest.

\section*{Acknowledgments}
Support come from the Key-Area Research and Development Program of Guangdong
Province under Grant No. 2019B030330001,  the National Natural Science Foundation of China
under Grant No. 11774436, No. 11974434 and No.12074439, the fundamental research funds for the Central Universities (Sun Yat-sen University, 2021qntd28). Le Luo receives support from Guangdong Province
Youth Talent Program under Grant No. 2017GC010656, Sun Yat-Sen University Core Technology Development Fund. Yang Liu receives support from Natural Science Foundation of Guangdong Province under Grant 2020A1515011159. Ji Bian receives support from China Postdoctoral Science Foundation under Grant 2021M703768. 


\bibliographystyle{unsrt}
\bibliography{anti_pt}

\end{document}